# ENVISION OF AN INTEGRATED INFORMATION SYSTEM FOR PROJECT-DRIVEN PRODUCTION IN CONSTRUCTION

Ricardo Antunes[1] and Mani Poshdar[2]

## ABSTRACT

Construction frequently appears at the bottom of productivity charts with decreasing indexes of productivity over the years. Lack of innovation and delayed adoption, informal processes or insufficient rigor and consistency in process execution, insufficient knowledge transfer from project to project, weak project monitoring, little cross-functional cooperation, little collaboration with suppliers, conservative company culture, and a shortage of young talent and people development are usual issues. Whereas work has been carried out on information technology and automation in construction their application is isolated without an interconnected information flow. This paper suggests a framework to address production issues on construction by implementing an integrated automatic supervisory control and data acquisition for management and operations. The system is divided into planning, monitoring, controlling, and executing groups clustering technologies to track both the project product and production. This research stands on the four pillars of manufacturing knowledge and lean production (production processes, production management, equipment/tool design, and automated systems and control). The framework offers benefits such as increased information flow, detection and prevention of overburdening equipment or labor (Muri - 無理) and production unevenness (Mura - 斑), reduction of waste (Muda - 無駄), evidential and continuous process standardization and improvement, reuse and abstraction of project information across endeavors.

## INTRODUCTION

In manufacturing, the operation is constantly monitored by the supervisory control and data acquisition (SCADA) system. The system monitors, gathers, and processes real-time data from devices such as sensors and cameras, recording events into a log file and/or displaying the operational information to local and/or remote locations through human-machine interface (HMI) software. Because the information is available as soon as possible corrective actions can be taken almost immediately. With the current advancements in computing, intelligent models can also run in real time to detect future issues supporting preventive actions. Despite SCADA systems and automation being standard production tools in manufacturing their use in construction is minimal and limited to isolated equipment.

---

[1] The University of Auckland, Auckland, New Zealand, +64 20 40 12 4793, rsan640@aucklanduni.ac.nz
[2] Lecturer, The Auckland University of Technology, Auckland, New Zealand, +64 921 9999 ext. 8956, mani.poshdar@aut.ac.nz

In manufacturing, the production moves from machine to machine, worker to worker, or a combination of both. The route of production is fixed (Antunes and Gonzalez 2015; Hayes and Wheelwright 1979). Thus, the positions of sensors and actuators are fixed and planned according to the production routes and its flow. Once set, the positions only need to be modified if the production routes change. In construction production routes are flexible. "Jobs arrive in different forms and require different tasks, and thus the equipment tends to be relatively general purpose (Hayes and Wheelwright 1979)." Some production routes will only exist long after the beginning of the project by the time that others would be extinct. Construction must then rely on general purpose sensors that, as the equipment, can be used in different applications through the project life-cycle, often, requiring those also to be mobile. Hence, traditional instrumentation (and sensor positioning) used in a manufacturing SCADA systems do not work in construction, as the instrumentation must be mobile.

Building Information Modeling (BIM) can be considered as the closest system to a SCADA applied in construction. BIM is the only system in construction that may contain the production layout. However, BIM focuses mostly on production planning (Nederveen and Tolman 1992; Rossini et al. 2017). The monitoring and control are still performed manually regardless of the use of BIM. The production aspect of BIM, as well the general industry, relies on primitive project management practices such as critical path and Gantt charts [the latter neither being the first nor the most sophisticated production tracking approach (Antunes 2017; Wesolowski 1978)]. These obsolete practices have been abandoned in the industries with high productivity, such as information technology. Construction occupies the bottom of productivity charts even showing negative indexes of productivity over the years (National Society of Professional Engineers 2014). Some common issues are lack of innovation and delayed adoption, informal processes or insufficient rigor and consistency in process execution, insufficient knowledge transfer from project to project, weak project monitoring, little cross-functional cooperation, little collaboration with suppliers, conservative company culture, and a shortage of young talent and people development (Almeida and Solas 2016).

Although much work has been done on implementing information technology and automation in construction their application on an integrated flow of information is sparse. This paper proposes a framework based on the current literature and technology to implement automatic monitoring and control for construction management and operations that could be useful to address the biggest issues of production in construction. Conjointly, this research uses four pillars of manufacturing knowledge and Lean production: production processes, production management, equipment/tool design, and automated systems and control. The goal should be achieved by both top-down and bottom-up approaches. The top-down approach will tackle the production system collecting information about the construction environment and its changes. The bottom-up approach will analyze the worker's activity. By using smart-tools, embedded hardware, Internet-of-things (IoT) and tracking the effort of labor can be measured and related to project progress. The two approaches are stitched together by a machine learning engine which makes sense of the data and the production theory comparing what has been done with the plan provided in the BIM model.

# TECHNOLOGY

## BUILDING INFORMATION MODELING

BIM is a powerful, yet 'promising' tool for the design and construction industries. 'Promising' standing for both what it can do at the present and in the future. BIM is still seen as a new technology in construction despite the increasing adoption and awareness of BIM over the years (McGraw Hill Construction 2012; National BIM 2017).

The concept of BIM can be pinpointed back to the year of 1962 when Engelbart presented a hypothetical description of computer-based augmentation system (Engelbart 1962). The application of computational solutions in construction was researched a bit later

(Eastman 1969, 1973). The research focused on the automated space planning using artificial intelligence in the bi-dimensional realm. The term 'Building Information Management' appeared 30 years later (Nederveen and Tolman 1992) while the first commercial implementation using this term is credited to ArchiCAD (successor of Radar CH from 1984 for the Apple Lisa Operating System). Historically, it is important to note that BIM did not derive from bi or tri-dimensional CAD. BIM (concept) is contemporary of CAD development. Nevertheless, BIM as a tool built upon CAD three-dimensional design tools for building modelers, which was a fully developed graphical tool for building modeling available at the time.

The manufacturing industry explored further benefits of the tool besides graphical modeling, in particular, parametric information technology tools (Autodesk 2002). Forms in CAD drawings evolved to objects with the development of object-oriented programming languages and their implementation to CAD systems in the early 1990's. Objects can bear graphical and non-graphical information bringing advances in both areas. From a graphical perspective, instead of drawing elements, one by one the user could design them separately and insert and reuse objects in the desired location. The group of lines, forms, and surfaces is interpreted as a three-dimensional geometrical model of the element it represents, for instance, a door or a window. The non-graphical perspective gives meaning to that object. The object contains multiple graphical information, such as the drawings of the door opened and closed. The object can have parts, and these parts can be of different materials with different properties. The objects may also contain production information attached, such as cost, labor, schedule, and effort what will give

BIM means to serve as a planning tool. Furthermore, changes required can be done in the element and automatically replicated where it has been used rather than laborious one by one changes. Overall, the reusable objects can bare more details and libraries of objects could be developed and shared.

## VIRTUAL REALITY AND AUGMENTED REALITY

Both virtual reality (VR) and augmented reality (AR) make use of 3D models to create a scene in which the user can freely observe and/or interact with the models. What set these technologies apart is how they use the background where the objects lie. Virtual reality fully immerses the user providing a background to the environment. The user has the perception of being physically present in a non-physical world. Conversely, augmented reality utilizes the real environment as the background to project the models. The user is partially immersed. Each one has different applications. Using 3D models, VR can display a fictional scenario, for example, a functioning underground subway station even before excavations begin. AR requires a background, thus, at least part of the station must be in place. That is due to the fact that AR requires the recognition and tracking of environment specific points for object placement. Both VR and AR are useful as HMI.

## LIGHT DETECTION AND RANGING

Light Detection and Ranging (LiDAR) is a remote sensing method, which uses light reflection to measure distances. The emitter shoots the light (ultraviolet, visible, or near infrared) which is reflected and then captured by the receptor. As the speed of light, $c$, is known, the time between emission and reception, $t$, is used to calculate the distance, $d$, from the emitter to the reflector and back to the sensor, i.e., $t=2d/c$. An Global Positioning System (GPS) receiver and an Inertial Measurement Unit (IMU) provide the absolute position and orientation of the sensor. Thus, it is possible to calculate the position coordinates of the reflective surface. One implementation of LiDAR consists of a vertical array of emitters mounted on a rotational plate creating a linear scan that sweeps the surroundings at each rotation. The result is a cloud of points which describes the environment around the sensor. Despite the fact that the cloud of points provides accurate measurement; the data does not identify objects. Basically, this cloud consists of $x$, $y$, and $z$ coordinates of each point. Making sense of what a group of points is often is a manual task. Another limitation of LiDAR scans is the 'shadowing.' Because the technology relies on reflection, it can make sense of lies behind a reflective surface or at the non-reflective surface, such as water. The shadowing effect can be eliminated by scanning the environment from different locations and thus overlapping cloud points [once the LiDAR scans are almost ever combined with Global Positioning System (GPS) and inertial measurement units].

LiDAR has been integrated to BIM aiming to identify defects (Wang et al. 2015). That happens by comparing LiDAR measurements against BIM model specifications. Deviations out of determined bounds are then identified as defects (*Muda* – Level II). A quadcopter (any other carrier is possible, such as an aquatic or terrestrial drone or even a backpack) inspects the site using LiDAR (inspection may also be considered *Muda* of over-processing given the idea that the task should be done correctly instead of being inspected for the approval). In this approach the defect flag rises without human interaction, it however does not characterize a real-time system. The first reason is that the defect will only be detected when (and if) the drone finds the issue, not at the time the defect occurred. The second is that real-time systems require a timely response to the event. A response out of the time-frame often results in catastrophic failure. The response to the identified defect is not time-dependent. A real-time system depends on both the logical result of the event and the physical instant features (Kopetz 2011). For instance, the quadcopter drone moves forwards when detects on obstruction in its trajectory (event). The trajectory correction (response) must happen in a timely manner otherwise the drone will crash.

## IMAGE

Image analysis can be an important tool in construction with several applications during the project life-cycle. A simple application can the defect inspection, where the inspector reports the non-conformities by taking pictures of the items out of specification and which will feed a punch list to be addressed by the contractors (*Muda* of rework). The pictures are used as evidence of the status of the non-conformities detected.

Additionally, because special cameras/lenses/sensors can capture infra-red and ultraviolet, which are invisible to the human eye, the collected information can be used for evaluating thermal and light insulation. Depending on what the cameras are mounted, they can provide visual information from specific angles that are known to be dangerous for human inspection (e.g., in confined spaces), or even impossible (e.g. for the pipelines). The combination of multiple images provides even more information. Aerial mapping, elevation level, and 3D mapping are some examples in which several images are stitched together. Moreover, the image analysis process can be repeated periodically what will result in the visual representation of the evolution of a particular area or item over time. Nevertheless, image do not supply accurate measurements to what they represent. To add accurate quantitative data to images, these can be combined with LiDAR (Fei et al. 2008) or with sequenced BIM models (Skibniewsk 2014).

## CHRONO-ANALYSIS

Chrono-analysis is the assessment of footages to evaluate production. The footages are captured by cameras positioned around the shop floor to record an activity done by the worker(s). The time spent by the worker(s) on each task of the activity can be measured by watching the recordings. The tasks are classified into three major categories: value-adding, non-value-adding but necessary (*Muda* – Level I), and non-value-adding and unnecessary (*Muda* – Level II). One of the mantras of Lean is "eliminate *Muda*." Accordingly, first, the analyst will identify each task that composes the activity. For instance, the footage of the activity contains the recordings of the set-up, the core task, breaks and the cleaning. Next, the analyst chronographs each task. Then, the analyst plan on how to eliminate or at least minimize the time spent on non-value-adding tasks (set-up, breaks, and cleaning). Then, the analyst plan on how to eliminate or at least minimize the time spent on non-value-adding tasks (set-up, breaks, and cleaning). Later, the analyst implements the plan and potential solutions. The analyst's records new footage of the activity execution and tracks the time spend on the tasks. After a comparison of the times to the original results, the analyst updates the activity standard with the solutions that resulted in improvement. Chrono-analysis can be seen as lean focused, more detailed, and evidential implementation of time and motion analysis. The *caveats*: chrono-analysis is usually a laborious process conducted eventually rather than continuously; the benefits for tasks with a low level of repetitiveness are minuscule.

## PRODUCTION THEORY

The traditional theory about fundamental mechanisms of production in repetitive processes in construction is at an embryonic stage and does not yet fully establish the foundations of a production model. The traditional and convenient approach to project-driven production in construction is to rely on linear steady state models. By considering the transient state, Productivity Function produces models that are more accurate in describing the processes dynamics than the steady state approaches (Antunes et al. 2017). The Productivity Function provides a mathematical foundation to develop algebraic for the calculations of cycle times (average, best- and worst-cases), throughput at capacity (Antunes et al. 2018), and the influence of the transient state time in the production variability (Antunes et al. 2016).

Productivity Function has been applied in feedback loop control yielding a controlling approach [Productivity Function Predictive Control (PFPC)] that can achieve high performances even when processes operate closer to capacity (Antunes 2017). Moreover, this performance enhancement is higher when PFPC is applied to processes in a parade-of-trades (Tommelein 1998). The PFPC shown to be a robust approach to plan, control, and optimize production and supply chain in construction with direct implications to management practices such as *takt* time. A benefit of PFPC is its focus on minimizing the variances of output to the set point or plan. The PFMPC can operate satisfactorily even without an accurate model (Antunes 2017). In practice, the use of adaptive PFPC (APFPC) can be useful. This adaptive version estimates a Productivity Function cyclically within a period; thus, the control relies on a model that is accurate to the current time frame. Therefore, if the production system evolves (which is the goal of continuous improvement) that makes the model obsolete, APFPC can relearn the process and estimate a new model automatically.

Although the Productivity Function can describe a variety of systems (including multi-variables systems), a structure that can embrace nonlinear and/or time-variant systems is required; and respectively, the introduction of linear time-varying space-state models which can also describe nonlinear systems. Nevertheless, the evaluation of these function from the data is based on the back-propagation algorithm (Antunes 2017), which is a machine learning tool.

## MACHINE LEARNING

Machine learning is the term used to describe a field in computer science where the machine is trained on how to do a task instead of being programmed. Thus, by being trained (or training itself) the machine can develop its own way of how to execute the task (Silver et al. 2017). The training can be either assisted or unassisted. Assisted training means that the inputs and outputs are provided to the machine that makes sense of the conditions to determine the output. For unassisted training, only the inputs are available. That entails enormous flexibility to machine learning and its applicability. As such, machine learning can mix a variety of input sources (features) to determine or classify the output, being capable of performing simple (such as an and operation) to complex tasks. For instance, it can evaluate labor processes as numerical values to estimate a non-linear productivity function (Antunes 2017), or identify and track different elements at once in a video feed (Gordon et al. 2017).

# FRAMEWORK

The top-down and bottom-up approaches interact joining theory and practice in continuous improvement loop. This suggestion stands on two tenets: observer effect and *Genchi Genbutsu*. In physics, the term observer effect (Bianchi 2013) defines the influence of the observation act to the event. It means that by observing an event, the observer may alter the event, and consequently modify the observation. This effect is also known in the human sciences, where subjects have their behavior affected by being observed. In this sense, the awareness of being observed may modify the production system and its model. Thus, production is constantly observed, and the information is used to modify production. *Genchi Genbutsu*, a principle of the Toyota Production System, which means 'go to the source and get the facts to make the right decision.' In this approach, instead of asking for information updates the progress status is obtained in real time from positioned sensors or upon inspection from the drones. Next, the machine learning engine will merge the information (LiDAR, images, sensors) with BIM to identify the product progress and deviations from the specifications (similarly as in the SCADA). In parallel, the production information (progress and workers information) is checked against the production theory and models to evaluate productivity, forecast conclusion dates and assess corrective actions (as in APFPC). These two combined and jointly with the project plan are then presented to the 'control room.' Therefore, the 'control room' can rely on accurate information in the decision-making process, which results in a data-driven continuous improvement loop (Figure 1).

For instance, if a fixed camera detects that a disposal bin is being filled at a certain rate the replacement of that bin can be ordered from the control room without the worker's requisition (that means eliminating the requisition task (*Muda* Level I), the work stoppage (*Muda* Level II) by waiting that the bin replacement or having to replace it (setup/cleaning, i.e., *Muda* Level I). And as *Muda* decreases *Mura* also decreases (Antunes et al. 2016). Similar reasoning works with suppliers. For example, if the casing were not cemented in place, the suppliers can be notified to avoid bringing more to the site. This integration with the supplier may avoid *Muda* (Level I) in one or more situations: inventory - use extra space to store more that its needed; waiting - if the trucks need to wait around the site; motion - case the truck needs to go back. Because information is compiled in the SCADA and centralized in the 'control room' it can be accessed and shared with ease, such as in a library.

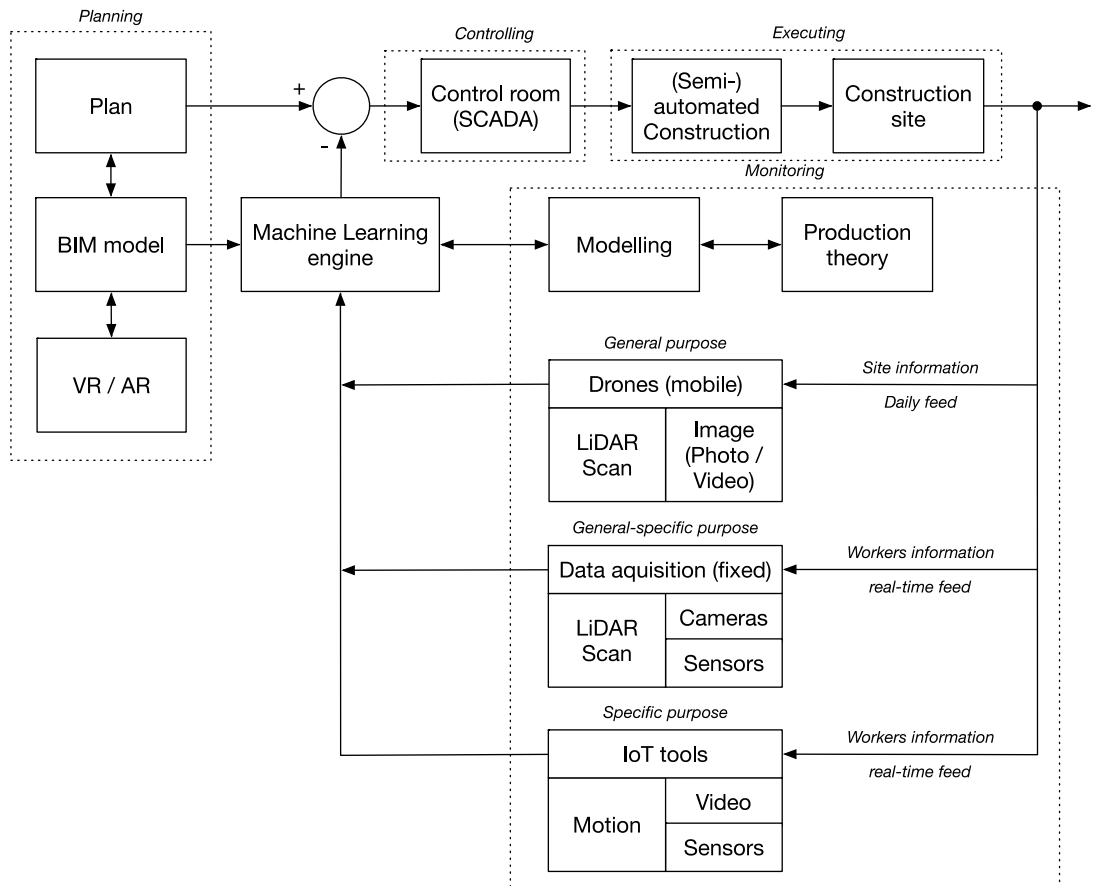

Figure 1 Theoretical framework for an information integration system for construction

## CONCLUSIONS

Cross-functional cooperation in construction is low mostly because the parts have no information about what is happening outside their area. The same can be said about suppliers. The establishment of the 'control room' centralizes information from the plan, labor, process, and production. Moreover, once the control-room has information about the progress of the current and next activities on-site, it will be able to coordinate cross-functional activities and supply chain.

Building information security and maintenance may use the product legacy information gathered in construction eliminating redundant work by analyzing the building. This work has already been done during construction (reducing over-processing). There is a compiled log of who did what, when and how for every part of the building including divergences between the original design and every change and defect occurred during construction. There can be extensive details of how the process has been done (and evolved). The production knowledge has further benefits. Especially, due to the network effect. The network effect adds value to this framework with use and adoption. This means that data can be generalized to a broader audience with more information such as season of the year, weather condition, geo-localization, altitude, winds, local culture, diversity, or any other feature. Hence, future endeavors will establish the production base-line using historical evidence rather than the usual labor/time relationship.

Using the chrono-analysis continuous assessment jointly with the data (production progress and workers effort) from previous projects informal processes tends to be eliminated. Better processes are developed and standardized. More accurate historical information is persistent and can be generalized to different projects enabling comparison and continuous improvement methodologies from project to project. New builders will be trained in the benchmark process instead of the "I have been done this for the last $x$ years" (and repeating the same mistakes over and over) approach. As such, the conservative company culture, lack of innovation and delayed adoption will be addressed by the marked. Companies will quantitatively assess and qualify the performance of contractors in previous projects. In an intensive third-party contracting industry such as construction, low productivity companies that often make mistakes are costly, and consequently, put at the end of the supplier's list or dismissed. Construction needs an increase in the number of builders, but it really needs builders with better performance.

A more automated construction industry should experience a set of benefits, such as better decision-making processes, increase information flow, and increase productivity. These benefits have a collateral impact on the whole society. More productivity means that more projects can be done using fewer resources. Accordingly, more infrastructure can be built and maintained. It can increase the affordability of the housing prices.